# A Novel Fault Detection Approach combining Adaptive Thresholding and Fuzzy Reasoning

I. Fliss

LI3 laboratory
University of Manouba, National
School of computer Sciences,
2010 Manouba, Tunisia.

Imtiez.Fliss@ ensi.rnu.tn

M. Tagina

LI3 laboratory
University of Manouba, National
School of computer Sciences,
2010 Manouba, Tunisia.

Moncef.Tagina@ ensi.rnu.tn

*Abstract--* **Fault detection methods have their pros and cons. Thus, it is possible that some methods can complement each other and offer consequently better diagnostic systems. The integration of various characteristics is a way to develop "hybrid" systems to overcome the limitations of individual strategies of each method. In this paper a novel detection module combining the use of adaptive threshold and fuzzy logic reasoning inspired by the Evsukoff's approach is proposed in order to reduce the rate of false alarms, guarantee more robustness to disturbances and assist the operator in making decisions. The proposed approach can be used in case of multiple faults detection. This approach is applied to a benchmark in diagnosis domain: the three-tank system. The results of the proposed detection module are then presented through a gradual palette of colors in the graphical interface of the system.**

## 1. INTRODUCTION

Fault detection is crucial in the process of system supervision. It consists in deciding if the physical process is faulty or not. Any undetected fault can reduce the system performance and have grave consequences. Due to modeling errors and unmeasurable disturbances, it is difficult to distinguish between the effects of an actual fault and those caused by uncertainties and disturbances.

To avoid instable decisions in these cases, many techniques have been used. Detection methods have their strengths and weaknesses. Thus, it would be interesting to integrate various techniques to overcome the limitations of individual strategies of each method and to ensure the reliability and effectiveness of the detection module.

In this context, our work consists in proposing a novel detection module which integrate the use of adaptive threshold and fuzzy logic reasoning inspired by the approach described in [9]. The proposed module can be used in case of multiple faults detection.

The results of the proposed detection module (which represents the state of the process) are shown through a gradual palette of colors in the graphical interface of the system from the green nominal state to the red faulty state. This leads to stable decisions and allows gradual information processing that can be easily understood by the human operator [11].

In this paper, the proposed fault detection module is introduced. Then, an application of our approach to a benchmark in diagnosis domain: the three-tank hydraulic system is presented.

The paper is organized as follows: the second section presents the state of the art of fault detection methods. While section 3 introduces the proposed fault detection module, section 4 is devoted to the presentation of our implementation. Finally, some concluding remarks are made.

## 2. FAULT DETECTION METHODS: THE STATE OF THE ART

Detection consists in deciding if the system is or not faulty regardless disturbances. In fact, the purpose of detection is to establish a rule of decision that can detect the earliest possible passage of a normal operating condition, called Hypothesis $H_0$, to a state of abnormal, where there are failures, noted Hypothesis $H_1$, corresponding to an unpredictable change of some parameters related to the process. When using such a decision rule, the operator wants to reduce the delay in detection and the rate of false alarms. To fully use the information contained in the process measures, detection consists of two phases: the generation and the evaluation of residuals.

### A. The residuals' generation

The generation phase is based on a model of the physical system to be monitored. This step consists in calculating the residuals which are consistency indicators between recorded measures and the model behavior. The residuals' generation is often based on quantitative models.

### B. The residuals' evaluation (decision)

The evaluation phase (converting the residuals' values in symptoms) is not simple since the model is never perfect and measures are usually marred by noise and/or uncertainties. In practical cases a residual is never zero.

Therefore, we must make a difference between "low" residuals which are characteristic of normal state and "big" residuals which indicates the presence of faults. We find several techniques to evaluate residuals:

*1) Thresholding:*

Evaluation, in this case, consists in defining a threshold to detect the presence of faults. The main difficulty of this method is the threshold's calculation. A too high threshold may lead to non-detection. Instead, a too small threshold will cause false alarms. The problem is to find a threshold which will be the best compromise between minimum false alarms and non-detection rates.

Early works focused on the development of fixed thresholds, independent of time and system's inputs. For instance, Walker and Gay [19] have defined fixed thresholds using the Markov theory (Markov chain).

Notice that a measure uncertainty can lead to a residual which excesses limits and triggers a false alarm. Thus, Emami-Naeini, Akhter and Rock [8] have defined the concept of adaptive threshold which is robust against the uncertainties of model. The adaptive threshold is, in this case, a compromise between the false alarms and bad detections. Clark follows the same approach improving the detection step with an adaptive threshold decision defined empirically using deterministic functions of system's inputs [16]. There are many contributions in the field of adaptive thresholds generation [13], [14]…

The advantage of the adaptive threshold is that it can guarantee a constant rate of false alarms throughout the process. The use of this kind of threshold also allows keeping control over the rate of false alarms and the likelihood of false diagnoses.

*2) Statistical decision:*

For this assessment, we can proceed with statistical methods which are (without being exhaustive):

- Generalized Likelihood Ratio (GLR) test introduced by Willsky and Jones [20] taking into account the stochastic perturbations (white noise test).
- Page-Hinkley test which is based on testing the residual average value on a detection window compared to a preset threshold. [3]
- Decorrelation filter which remove, under certain hypotheses, the influence of model uncertainty on the residual. [5]

The advantage of statistical methods is that they can usually evaluate the probability of false alarms and non-detection. However, these methods require failure modeling knowledge and can't detect new faults (never seen).

The drawback of these methods is that the false alarms rate increases with the lower rate of fault detection and detection delay. Another difficulty with this method is the large amount of data needed for the quantities estimation such as statistical average or standard deviation. It is also physically difficult to interpret the detection result.

*3) Fuzzy Decision:*

Fuzzy decision methods can be used to evaluate residuals, considering that the concept of "zero residual" is a vague concept that can be described as a fuzzy set [10]. Fuzzy logic is considered in [6] as the best framework in which we can handle uncertainties and which also allows the treatment of some incompleteness. This approach allows taking into account the size of the residual, its persistence in time. This approach is widely used in the literature: [9], [11], [12] …

*4) Modal Interval Decision:*

Another way to evaluate residuals is to use the interval calculation. It is assumed that we have an interval to characterize unknown model parameters, noise measurement. These intervals represent the values considered acceptable. We deduce then the system outputs interval [2]. This calculation is not always easy in dynamic case. By comparing this interval to the measured interval it can be concluded, if they have no intersection, that there are faults. A study done by Heim [12] showed that decision is delayed and there are some gaps in the detection.

As a conclusion, we notice that the methods of adaptive threshold and the fuzzy decision seem to be more advantageous compared to other techniques. Thus, we decided to retain these two methods and incorporate them into the diagnostic system we propose in order to ensure the reliability and effectiveness of the detection module to be developed.

## 3. THE PROPOSED FAULT DETECTION APPROACH

It is obvious that detection methods have their pros and cons. Thus, it is possible that some methods can complement each other and offer consequently better diagnostic systems. The integration of various characteristics is a way to develop "hybrid" systems to overcome the limitations of individual strategies of each method. Therefore, it would be interesting to combine decision methods to improve detection phase and give representative information for decision-making to the operator. Our detection approach consists in combining the use of adaptive threshold and fuzzy approach inspired from Evsukoff and al' work [9] in order to reduce the rate of false alarms, guarantee more robustness to disturbances, assist the operator in making decisions and detect multiple faults.

Fig. 1 describes the proposed fault detection module.

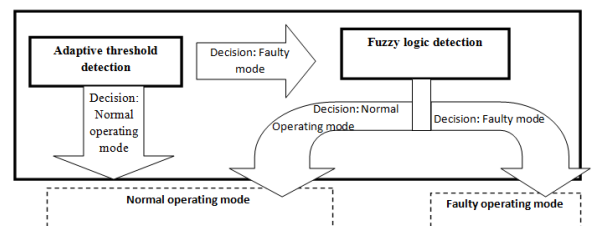

Fig. 1. The proposed fault detection module

The first step of our detection module is the use of adaptive threshold:

### A. The used adaptive threshold fault detection method

The use of an adaptive threshold, instead of a fixed one, can improve the performance of a fault-detection module significantly with respect to delay of detection and false alarm rate [13].

In our work, we rely on the threshold described in [13], which consists of a filter, with lead-lag behavior driven by the input signal (see Fig. 2). This causes the filter output to be zero for steady-state inputs.

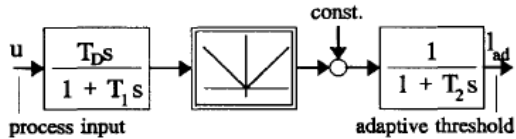

Fig. 2. Generation of an adaptive threshold [13]

The complete structure contains four parameters which must be determined: A constant factor 4/3 times the maximum residual value in the fault free case with constant input signals is used. Time constants $T_1$ and $T_2$ are chosen according to the cut-off frequency of the state variable filter represented in the Fig. 3, respectively. The ratio $\dfrac{T_D}{T_1}$ is a measure for the uncertainty of the dynamics.

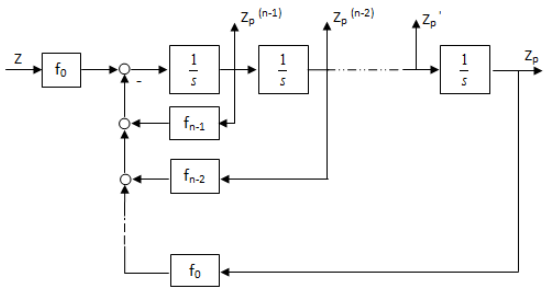

Fig. 3. State variable filtering [13]

Where fi are Butterworth-filter with an adequate cut-off frequency wc. Wc is chosen so that it is close to the highest process eigen-frequency which guarantees that the important information of the signal is kept and higher frequencies according to disturbances are well suppressed [13]. The use of adaptive threshold ensures a constant and low false alarms rate throughout the process [4]. However, this technique is not too robust to disturbances.

To improve the results given by this method, more reduce the rate of false alarms and guarantee more robustness to disturbances; it would be interesting to integrate in our detection module an increasing considerably used method: fuzzy logic. Indeed, this method can take into account the uncertainties by the gradual nature of belonging to a fuzzy set.

### B. The used fuzzy logic detection method

Fuzzy logic fault detection consists in interpreting the residuals by generating a value of belonging to the class AL (alarm) between 0 and 1 that allows to decide whether the measurement is normal or not. The gradual evolution of this variable from 0 to 1 represents the evolution of the variable to an abnormal state. In our work, this evolution is then changed into an index characterizing a color code (green, red) to represent the state of different variables on a causal graph.

Before starting the first step of fuzzy logic procedure which is the step of fuzzification, the inputs, outputs and the fuzzy partitions have to be defined.

Thus, we set the analytical redundancy relations to be the input variables of the fuzzy logic detection module. The outputs are the variables' states. The linguistic input variables are defined as {NB, N, Z, P, PB} meaning negative big, negative, zero, positive and positive big respectively.

The membership functions of the inputs are shown in Fig. 4. The membership input functions' parameters are -a4, -a3, -a2, -a1, a1, a2, a3 and a4.

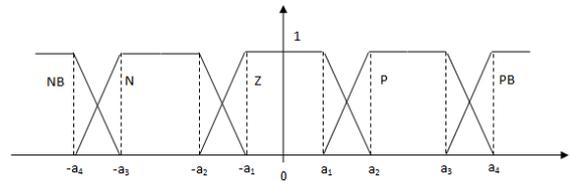

Fig. 4. Input Fuzzy partitions

Symmetrical trapezoidal membership functions are used for fuzzy partitions. Symmetry is justified here by the fact that in the absence of a fault, residuals are zero, and positive or negative faults have the same importance [9]. The symmetry leads to a simple parameterization of the residual value fuzzy partition with only four parameters: a1, a2, a3 and a4 (instead of eight) corresponding to the trapezoid boundaries.

On the other hand, an output can be OK or AL. Fig. 5 shows the different outputs' membership functions. Note that the membership output functions' parameters are a, b, c and d.

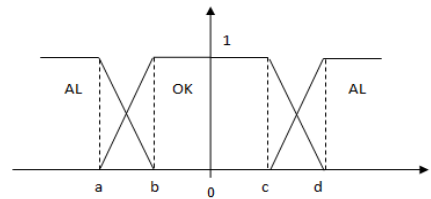

Fig. 5. Output Fuzzy Partitions

Each conclusion is associated with a color. The description obtained as an output of fuzzy reasoning in detection module is defuzzified into an index associated with a color shade to represent how each of the conclusions is matched. In our case, the conclusions are defuzzified into an index of a color map from green (normal state) to red (faulty state).

Colors codes have been widely used in human-machine interfaces to represent qualitative numerical results. Using colors enables concepts that gradually evolve with time to be represented. A graduated color code translates the detection algorithm results into a mass visualization of the variable states, which allows operators to analyze a large amount of qualitative information [9].

## 4. EXPERIMENTAL RESULTS

To test the performance of the proposed approach, we have chosen a benchmark frequently used in diagnosis domain: the three-tank hydraulic system [1] [15] [17] …

### A. System presentation

Fig. 6 illustrates the three-tank system. The process consists of three cylindrical tanks. Tanks communicate through feeding valves.

The process has two inputs: Msf1 and Msf2. We put five sensors: effort sensors De1, De2 and De3 to measure pressure of C1, C2 and C3 and flow sensors Df1 and Df2 measuring flow level of the valves 1 and 2. Its global purpose is to keep a steady fluid level in the tanks.

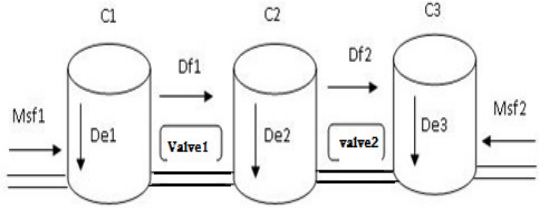

Fig. 6. The Three-tank system

In our work, we were based on Bond Graph modelling [7] to get the analytical redundancy relations.

Fig.7 presents the Bond Graph model of the three-tank system.

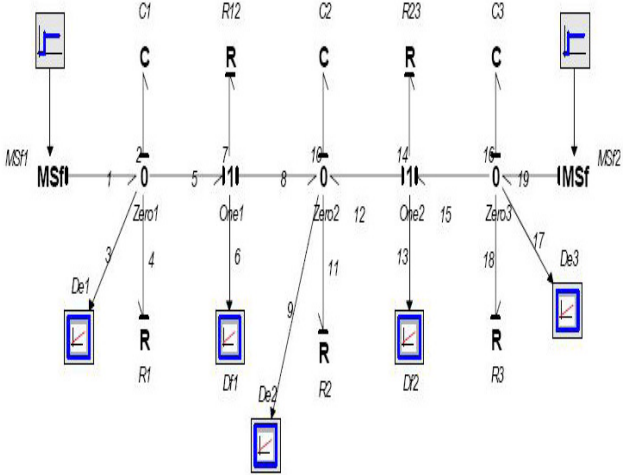

Fig. 7. Bond Graph model of the three-tank system.

Using the procedure described in [18], we obtain the following analytical redundancy relations (RRAi, i∈ {1, 2, 3, 4 and 5})):

RRA1:
$$\frac{1}{sC_1} MSf_1 - (1 + \frac{1}{R_1 C_1 s}) De_1 - \frac{1}{C_1 s} Df_1 = 0 \quad (1)$$

RRA2:
$$\frac{1}{sC_2} Df_1 - (1 + \frac{1}{R_2 C_2 s}) De_2 - \frac{1}{C_2 s} Df_2 = 0 \quad (2)$$

RRA3:
$$\frac{1}{C_3 s} MSf_2 - \frac{1}{C_3 s} Df_2 - (1 + \frac{1}{R_3 C_3 s}) De_3 = 0 \quad (3)$$

RRA4:
$$\frac{De_3 - De_2}{R_{23}} - Df_2 = 0 \quad (4)$$

RRA5:
$$\frac{De_1 - De_2}{R_{12}} - Df_1 = 0 \quad (5)$$

To test the performance of the proposed fault detection module, we inject faults at Msf1, Msf2 and De2.

In the following parts, we expose the results we get in this case.

Note that we consider in our work additional faults modeled as additive signals added to the variables of the three-tank system.

The first step of the proposed detection module is using adaptive threshold detection whose results are presented in the following part:

### B. Results of Adaptive threshold detection

The following figures show examples of some analytical redundancy relations (rrai, i∈ {1, 2 and 4}) responses and the corresponding adaptive thresholds we get.

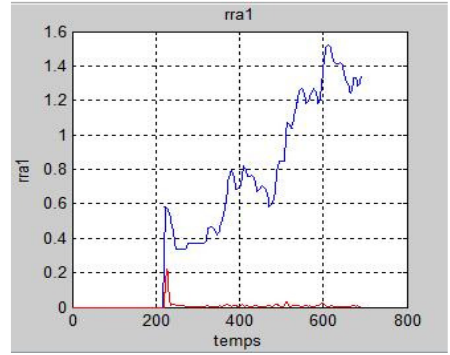

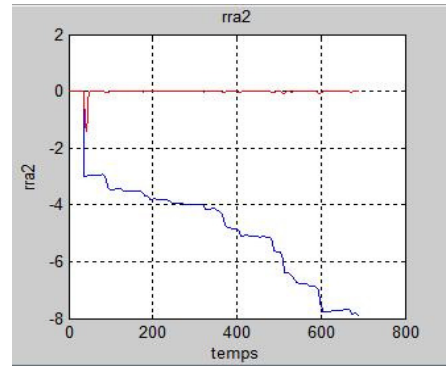

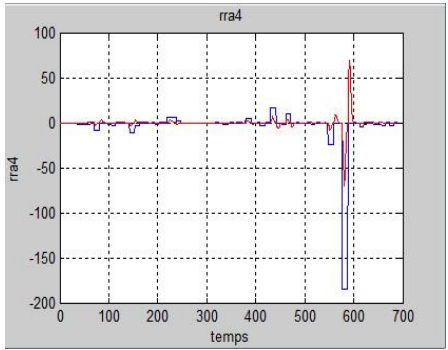

Fig. 8. Examples of rrai responses and their adaptive thresholds

According to this technique, the process is faulty. But in order to check the process state and reduce the risk of false alarms we execute the second phase of the proposed approach: the fuzzy fault detection inspired from Evsukoff et al's works [9].

*C. Fuzzy logic detection*

We proceed by fuzzy reasoning to determine the status of the system: normal or faulty displaying the results as a causal graph representing the state of the different variables through a gradual palette of colors from the green nominal state to the red faulty state.

Fig. 9 shows the result given by this method in the case we inject faults to variables Msf1, Msf2 and De2 of our three-tank-system: the red variables are suspected to be faulty; those in green have normal behavior. This enhances that the system is faulty.

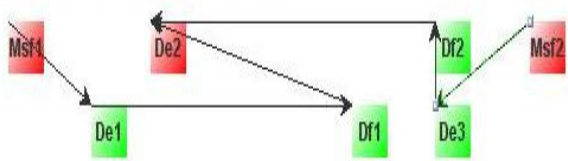

Fig. 9. Results of the proposed fault detection module

Our detection module finally announces that the process is actually faulty.

This proves that this method gives the right decision: the system is faulty. We notice that there is no false alarm of the adaptive threshold method in this case.

*D. Experimental results*

Two detection modules, one using only adaptive threshold, the other using the proposed approach, are adopted so that we can compare their efficiency in the procedure of fault detection and their rate of false alarms in case of multiple faults.

Several independent experiments are executed (a twenty tests are maid).

A comparison between the experimental results given by fault detection using only adaptive threshold and those given by our detection module is maid and shown in the table I and Fig. 10.

TABLE I

Experimental results

| % of false alarms using only adaptive threshold detection | % of false alarms using our detection module |
|---|---|
| 30% | 5% |

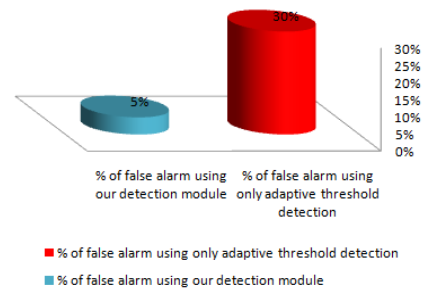

Fig. 10. Experimental results

Results provide evidence that the fault detection module proposed gives the proper decisions. It is effective in the procedure of fault detection. It is also clear that the use of the proposed detection approach presents an average reduction of 83.3% of false alarms given in the case of the use of the adaptive threshold only.

This can be explained by the fact that the integration of the two detection techniques overcomes the limitations of individual strategies of each method.

To conclude, this approach led to reduce the rate of false alarms, guaranteed more robustness to disturbances and assisted the operator in making decisions through the use of causal graph for the description of process variables' state. It proves also its efficiency in case of multiple faults detection. This technique presents a great promise for detection process.

## 5. CONCLUSION

In this paper, a novel detection module is proposed. The proposed solution consists in integrating adaptive threshold detection and fuzzy logic reasoning. In this way we could benefit from pros of both adaptive threshold and fuzzy logic methods in order to reduce the rate of false alarms and get more efficient diagnosis.

The application of this method to the three-tank system proves that the proposed detection module gives very good results compared to the detection module based only on the use of adaptive thresholding.

It is expected that the achieved results can also be extended to detect faults in genuine systems. In fact, we intend in future works to highlight the potential of using such a method in real application.